# STATISTICAL MODELING OF THE TIME COURSE OF TANTRUM ANGER

By Peihua Qiu[1], Rong Yang and Michael Potegal[1,2]

*University of Minnesota, Bristol-Myers Squibb and University of Minnesota*

Although anger is an important emotion that underlies much overt aggression at great social cost, little is known about how to quantify anger or to specify the relationship between anger and the overt behaviors that express it. This paper proposes a novel statistical model which provides both a metric for the intensity of anger and an approach to determining the quantitative relationship between anger intensity and the specific behaviors that it controls. From observed angry behaviors, we reconstruct the time course of the latent anger intensity and the linkage between anger intensity and the probability of each angry behavior. The data on which this analysis is based consist of observed tantrums had by 296 children in the Madison WI area during the period 1994–1996. For each tantrum, eight angry behaviors were recorded as occurring or not within each consecutive 30-second unit. So, the data can be characterized as a multivariate, binary, longitudinal (MBL) dataset with a latent variable (anger intensity) involved. Data such as these are common in biomedical, psychological and other areas of the medical and social sciences. Thus, the proposed modeling approach has broad applications.

**1. Introduction.** Anger is an important emotion that can intrude into daily life as low intensity irritation. At higher intensity, it underlies overt aggression at great social cost. For example, it has long been established that anger can trigger partner abuse and assault [e.g., Burgess et al. (2001),

Received January 2008; revised December 2008.

[1]Supported by grants from the National Institute of Child Health and Human Development (R21 HD048426) and from the Viking Children's Fund.

[2]Supported by a grants from the National Institute for Mental Health (R03-MH58739), from the Harry Frank Guggenheim Foundation and by National Research Service Awards from the National Institute for Neurological Disorders and Stroke (F33 NS09638) and the National Institute of Child Health and Human Development (F33 HD08208).

Jacobson et al. (1994), Schumacher et al. (2001)]. More recently, it has been found to play a role in incidents of road rage [e.g., Lupton (2002), Parker, Lajunen and Summala (2002)]. Anger at its most intense has been claimed to play a causal role in 35% of homicides [Curtis (1974)]. To approach this socially important but poorly understood phenomenon scientifically, we need to be able to measure and quantify anger intensity. Unfortunately, little is known about quantifying anger [cf., Fridja et al. (1992)]. The self-ratings of anger intensity on subjective rating scales (e.g., 1–10) used in classic emotion research are not open to verification. In contrast, people's behavior can reliably indicate the intensity of their anger. For instance, a grimace or a grunt suggests irritation, a shout indicates anger, and a screaming assault demonstrates rage. According to Sonnemans and Frijda (1994), the severity of angry action is one of the strongest predictors of the overall intensity of felt anger. Thus, an alternative model of anger would involve quantifying anger intensity based on a set of observable behaviors.

In a typical episode, anger first rises and then falls. What little is known about this trajectory from self-report studies suggests that anger intensity rises rapidly and then declines slowly [e.g., Beck and Fernandez (1998), Fridja et al. (1991), Tsytsarev and Grodnitzky (1995)]. Characterizing this trajectory is essential in understanding how different angry behaviors are distributed within an episode of anger. However, anger episodes vary widely in duration [see Potegal, Kosorok and Davidson (1996) for a review]. One resulting complication is that the trajectory of anger might vary systematically with duration. Thus, a complete model must specify not only the relationship between underlying anger and the overt angry behaviors, but also how the trajectory of anger varies as a function of episode duration. These are the major goals of this paper.

A problem in behavior-based modeling is that angry behaviors in adults are highly idiosyncratic and therefore difficult to compare or collapse across subjects. Furthermore, adults tend to mask their emotions in public situations and their angry behaviors are correspondingly difficult to observe. In contrast, the high frequency of young children's tantrums indicates that they tend not to mask their emotions [Underwood, Coie and Herbsman (1992)]. Also, the angry behaviors appearing in tantrums are stereotyped and similar across children. They are easily observed and identified by parents. For these reasons, our study focuses on young children's tantrums.

Earlier work suggests that tantrum behaviors, like the angry acts of adults, are differentially associated with low and high intensities of anger. For instance, stamping is associated with lower anger intensity, and screaming and shouting reflect higher anger intensity [Potegal and Davidson (2003), Potegal, Kosorok and Davidson (2003)]. In this context, it is of considerable interest that higher intensity anger behaviors are most likely to occur relatively early in tantrums, while lower anger behaviors are more evenly distributed [Potegal, Kosorok and Davidson (2003)]. These observations could be confirmed if the probability of lower anger behaviors did not change much with anger intensity, but the probability of higher anger behaviors were to increase strongly with anger intensity. Our analysis will determine if this is the case.

The data forming the basis of this analysis are derived from written parental narratives of tantrums of 296 children ranging from 18 to 60 months old, which were collected in the Madison WI area during the period 1994–1996 by Potegal and colleagues [Potegal and Davidson (2003), Potegal, Kosorok and Davidson (2003)]. Parents described each of their child's tantrums in as much detail as possible, including the durations of individual events or sets of events, within the tantrum. Detailed written instructions, checklists and examples guided parents in writing their narratives. Coders then converted the written narratives into time versus behavior matrices in which time was partitioned into consecutive 30-second units and eight different angry behaviors were scored as occurring or not within each unit. These behaviors are stiffen limbs/arch back, shout, scream, stamp, push/pull, hit, kick and throw. They are denoted as $x_1, x_2, \ldots, x_8$, respectively, in this paper. A tantrum begins with the first occurrence of one of these behaviors and it is over when all these behaviors disappear. This tantrum dataset can be characterized as a multivariate, binary, longitudinal (MBL) data associated with a latent variable, anger intensity. The observed angry behavior variables $x_1, x_2, \ldots, x_8$ are assumed to be driven by the (unobservable) anger intensity. Several indicators of data reliability, reported in Potegal and Davidson (2003), suggest that this dataset is reasonably reliable.

To model MBL data, generalized linear modeling [McCullagh and Nelder (1989)] and generalized estimating equations (GEEs) procedures [Diggle, Liang and Zeger (1994), Liang and Zeger (1986)] are natural options. Conventional GEE procedures would assume a generalized linear model for mean responses, with anger duration, time, etc. as predictors. Several authors, including Lin and Carroll (2001) and Severini and Staniswalis (1994), generalized the GEE method by including a nonparametric component in the mean responses in addition to a parametric component. However, these existing methods may not be appropriate for the current data because they do not allow us to study the relationship between the observed angry behaviors and the unobservable anger intensity. Wu and Zhang (2002) suggested a nonparametric procedure for modeling longitudinal data. But that procedure is for cases with univariate continuous responses, and it does not allow any latent predictors.

In this paper, a statistical modeling approach is suggested for describing the tantrum data. Our proposed model has three major components. First, we assume that the latent anger intensity, here called momentary anger (MA), follows a flexible parametric form over the time course of a tantrum. Although certain existing psychological research suggests an asymmetrical trajectory of MA with a rapid rise and slow decline, as mentioned above, our model must be capable of representing a range of possible trajectories. These include a high initial value (i.e., the anger intensity rises so rapidly that it appears instantaneous under our conditions of observation) followed by a slow decline over the entire tantrum, a symmetrical rise and fall, or a gradual increase over the entire tantrum. After an extensive comparison of various candidate functions, including polynomial functions and the Gamma function, the two-parameter Beta function is chosen for modeling MA (see Section 2 for a detailed description). Second, potential dependence of MA on tantrum duration is handled by allowing each of the two parameters of the Beta function to be a polynomial function of duration. Third, the relationship between MA and the likelihood of each of the eight angry behaviors is assumed to follow a generalized polynomial model. Model parameters are estimated by a proposed iterative algorithm, which is a modified version of the conventional GEE algorithm. In that algorithm, all model parameters are grouped into several blocks to speed up computation.

MBL data with one or more latent variables involved are quite popular in biomedical, psychological and other areas of medical and social sciences. For instance, observable sleeping status (i.e., deep sleep, light sleep or awake) of animals is believed to be affected by the unobservable circadian rhythm [cf., Qiu (2002), Qiu et al. (1999)]. The proposed modeling approach has broad applications in these and many other studies.

The remainder of this article is organized as follows: In Section 2 our proposed model for describing the tantrum data is described in detail. In Section 3 we apply this approach to the tantrum data and report some results. Several remarks conclude the article in Section 4. Identifiability of the proposed model and model estimation are discussed in the Appendices.

**2. Proposed model.** The tantrum dataset studied in this paper consists of 296 tantrums of widely varying duration. The shortest duration is only 0.5 minutes, while the longest tantrum lasts for 39.5 minutes. All 296 durations are summarized and displayed in Figure 1. In this plot, the durations are classified into the following six categories: 0.5–2 minutes, 2.5–4 minutes, 4.5–10.5 minutes, 11–20 minutes, 20.5–30 minutes, and 30.5–39.5 minutes. For convenience in plotting, these categories are slightly modified to the class intervals: $(0.25, 2.25]$, $(2.25, 4.25]$, $(4.25, 10.75]$, $(10.75, 20.25]$, $(20.25, 30.25]$, and $(30.25, 39.75]$. The densities of these class intervals, which are defined by the relative frequencies of durations in the intervals divided by the corresponding lengths of the intervals, are displayed in this plot as a density histogram. So, the area of each bar denotes the probability that a randomly chosen duration would be in the corresponding interval. In this plot, the integer above each bar denotes the frequency of the durations that are included in the corresponding interval. It can be seen that only about 10% of the tantrums are longer than 11 minutes.

One major goal of this research project is to describe temporal variation of MA and the relationship between MA and the eight angry behaviors in an entire tantrum episode. To this end, we first standardize observation times of angry behaviors along a 0 to 1.0 time scale. That is, the standardized time, denoted as $t$, for an observation in a given tantrum is the actual time of that observation divided by the tantrum duration. In other words, the standardized observation time $t$ is the percentile of the duration of that tantrum episode at the given observation time. Let $\pi_k(t)$ denote the probability of the $k$th angry behavior at time $t$, for $k = 1, 2, \ldots, 8$, and $MA(t)$ be the latent variable MA at $t$. Then, in this section, we discuss statistical modeling of $MA(t)$ and the linkage between $\pi_k(t)$ and $MA(t)$.

As noted above, some existing psychological research suggests that $MA(t)$ may increase rapidly at the beginning of the tantrum and then decline gradually. However, to avoid introducing a premature bias into determining the trajectory of MA, candidate functions must be able to assume a range of possible shapes. As mentioned above, after searching quite extensively among many commonly used parametric families of curves defined on $[0, 1]$ that could be used for describing $MA(t)$, we found that the following two-parameter family of Beta curves can serve this purpose well:

$$MA(t, r, s) = t^{r-1}(1-t)^{s-1} \qquad \text{for } t \in [0, 1],$$

where $r, s > 0$ are two parameters. Figure 2 presents several Beta curves. It can be seen that the two parameters of the Beta function make it highly polymorphic. By changing values of these two parameters, it can assume many different shapes, including the three possible shapes of $\mathrm{MA}(t)$ mentioned in Section 1.

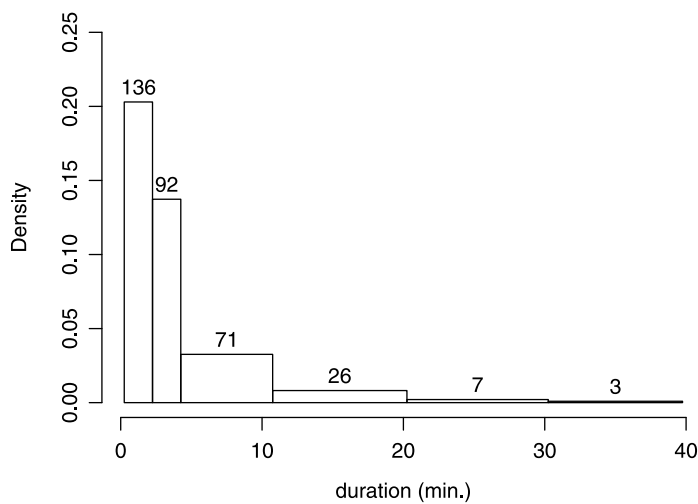

FIG. 1. *Density histogram of the durations of the tantrum dataset.*

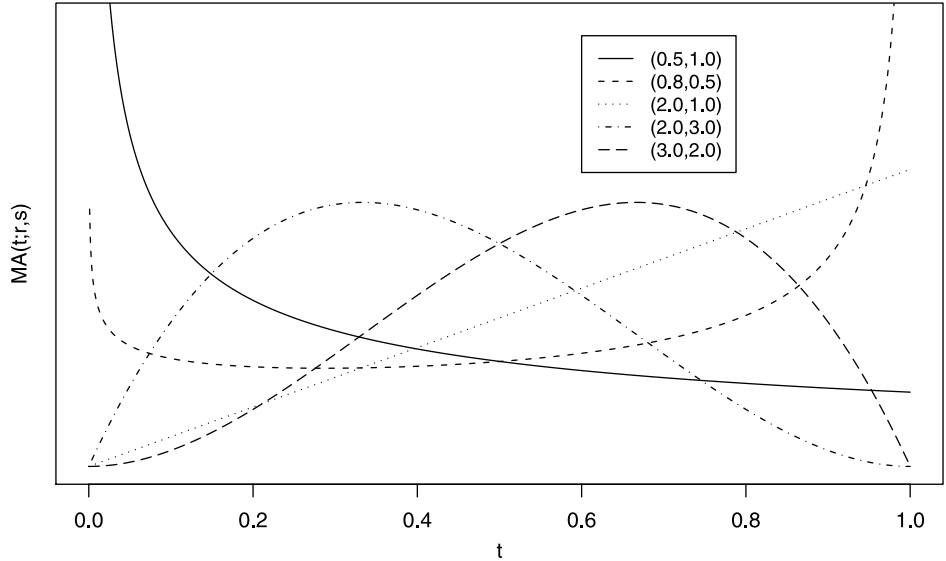

Fig. 2.  *Beta curves when parameters $(r, s)$ take the values listed in the legend box.*

In the above expression of $\mathrm{MA}(t, r, s)$, the restriction that $r$ and $s$ are positive would be inconvenient for model estimation discussed in Appendix B. For this reason, we reparameterize that expression by replacing $r$ by $e^a$ and $s$ by $e^b$, where $a$ and $b$ are two new parameters taking arbitrary values on the entire line $R$. After this reparameterization, $\mathrm{MA}(t, r, s)$ becomes

$$(1) \qquad \mathrm{MA}(t, a, b) = t^{e^a - 1}(1 - t)^{e^b - 1} \qquad \text{for } t \in [0, 1].$$

For simplicity of presentation, sometimes we write $\mathrm{MA}(t, a, b)$ as $\mathrm{MA}(t)$, which should not cause confusion.

To allow possible dependence of $\mathrm{MA}(t, a, b)$ on the tantrum duration, denoted as $d$, the two parameters $a$ and $b$ are assumed to be the following polynomials of $d$:

$$(2) \qquad a = a_0 + a_1 d + \cdots + a_{m_a} d^{m_a}, \qquad b = b_0 + b_1 d + \cdots + b_{m_b} d^{m_b},$$

where $m_a$ and $m_b$ are two non-negative integers. As shown in Figure 1, tantrum durations can change quite dramatically. In the psychological literature, researchers doubt that the trajectory of $\mathrm{MA}(t)$ may depend on tantrum duration [cf., Fridja et al. (1992), Potegal, Kosorok and Davidson (1996)]. Models (1) and (2) have the flexibility to allow such dependence.

As mentioned in Section 1, the eight observable angry behavior variables $x_1, x_2, \ldots, x_8$ are assumed to be driven by MA, each with its own linkage function. To model the linkage between $x_1, x_2, \ldots, x_8$ and MA, the following generalized linear model is used: for $k = 1, 2, \ldots, 8$,

$$(3) \quad \operatorname{logit}(\pi_k(t, \mathbf{a}, \mathbf{b}, \mathbf{c}_k)) = c_{0k} + c_{1k} \operatorname{MA}(t, a, b) + \cdots + c_{m_k k} \operatorname{MA}^{m_k}(t, a, b),$$

where $\operatorname{logit}(\pi_k(t, \mathbf{a}, \mathbf{b}, \mathbf{c}_k)) = \log(\pi_k(t, \mathbf{a}, \mathbf{b}, \mathbf{c}_k)/(1 - \pi_k(t, \mathbf{a}, \mathbf{b}, \mathbf{c}_k)))$, $\mathbf{a} = (a_0, a_1, \ldots, a_{m_a})'$, $\mathbf{b} = (b_0, b_1, \ldots, b_{m_b})'$, $\mathbf{c}_k = (c_{0k}, c_{1k}, \ldots, c_{m_k k})'$, and $m_k$ are nonnegative integers. In the above expression, we have made the coefficients $\mathbf{a}, \mathbf{b}$ and $\mathbf{c}_k$ explicit in the notation of $\pi_k(t, \mathbf{a}, \mathbf{b}, \mathbf{c}_k)$, for convenience of discussion about model estimation. Model (3) allows the log-odds of the likelihood of each behavior variable to be a polynomial function of $\operatorname{MA}(t, a, b)$. In Appendix A, we show that parameters in models (1)–(3) are all identifiable.

**3. Analysis of the tantrum data.** This section is organized in two parts. We first present a simulation study in Section 3.1 about the model estimation algorithm used in analyzing the tantrum data. Then, some results about the tantrum data are presented in Section 3.2.

3.1. *A simulation study.* The proposed models (1)–(3) are estimated by a modified version of the generalized estimating equations (GEE) algorithm. In that modified version, instead of updating all parameters of the models $\tilde{\mathbf{c}} = (\mathbf{a}', \mathbf{b}', \mathbf{c}_1', \ldots, \mathbf{c}_8')'$ simultaneously in each iteration of the GEE algorithm, we suggest dividing $\tilde{\mathbf{c}}$ into $g$ blocks, that is, $\tilde{\mathbf{c}} = (\tilde{\mathbf{c}}_1', \tilde{\mathbf{c}}_2', \ldots, \tilde{\mathbf{c}}_g')'$, and then updating the parameters block-by-block in each iteration of the algorithm. By doing so, computation can be greatly reduced. Details of the GEE algorithm and our proposed version with blocked parameters are described in Appendix B.

To check potential impact of the blocking scheme on parameter estimation by our modified version of the GEE algorithm [cf., expression (15) in Appendix B], we perform the following simulation study. It is assumed that there are $N = 200$ subjects in the study. For the $i$th subject, 3 binary variables are observed at $n_i$ equally spaced time points $t_{ij} \in [0, 1]$, where $j = 1, 2, \ldots, n_i$ and $i = 1, 2, \ldots, N$. Therefore, durations of all tantrums are assumed to be 1 in this example. It is further assumed that $n_i = 5$ when $1 \le i \le 50$, $n_i = 6$ when $51 \le i \le 100$, $n_i = 7$ when $101 \le i \le 150$, and $n_i = 8$ when $151 \le i \le 200$. Probability of occurrence of the $k$th binary variable at time $t_{ij}$ is assumed to follow the model

$$(4) \quad \operatorname{logit}(\pi_k(t_{ij})) = c_{0k} + c_{1k}(t_{ij} + \beta t_{ij}^2) \qquad \text{for } k = 1, 2, 3,$$

where true values of parameters are set to be $c_{01} = 0.5, c_{02} = 0.5, c_{03} = 0, c_{11} = c_{12} = c_{13} = 1$, and $\beta = -1$. Obviously, model (4) is a special case of models (1)–(3) when $e^a = e^b = 2$ in (1) and $m_k = 1$ in (3) for all $k = 1, 2, 3$.

TABLE 1
*This table presents the averaged estimates and the corresponding standard errors (in parentheses) of the parameters of model (4) by algorithm (15) in Appendix B with the three blocking schemes B-I, B-II and B-III. The results are based on 100 replicated simulations*

| Parameters | True values | B-I | B-II | B-III |
|---|---|---|---|---|
| $c_{01}$ | 0.5 | 0.5273 (0.0185) | 0.5279 (0.0185) | 0.5279 (0.0185) |
| $c_{11}$ | 1 | 0.8770 (0.0980) | 0.8694 (0.0981) | 0.8694 (0.0981) |
| $c_{02}$ | 0.5 | 0.5105 (0.0157) | 0.5132 (0.0159) | 0.5132 (0.0159) |
| $c_{12}$ | 1 | 0.9287 (0.0869) | 0.9168 (0.0878) | 0.9168 (0.0878) |
| $c_{03}$ | 0 | $-0.0074$ (0.0156) | $-0.0101$ (0.0153) | $-0.0101$ (0.0153) |
| $c_{13}$ | 1 | 1.1041 (0.0788) | 1.1152 (0.0773) | 1.1152 (0.0773) |
| $\beta$ | $-1$ | $-1.0170$ (0.0224) | $-1.0061$ (0.0211) | $-1.0061$ (0.0211) |

In the modified GEE algorithm (15), we consider the following three blocking schemes:

B-I: $g = 1$ and $\tilde{\mathbf{c}}_1 = (c_{01}, c_{11}, c_{02}, c_{12}, c_{03}, c_{13}, \beta)'$,

B-II: $g = 2$, $\tilde{\mathbf{c}}_1 = (c_{01}, c_{11}, c_{02}, c_{12}, c_{03}, c_{13})'$, and $\tilde{\mathbf{c}}_2 = \beta$,

B-III: $g = 4$, $\tilde{\mathbf{c}}_1 = (c_{01}, c_{11})'$, $\tilde{\mathbf{c}}_2 = (c_{02}, c_{12})'$, $\tilde{\mathbf{c}}_3 = (c_{03}, c_{13})'$, and $\tilde{\mathbf{c}}_4 = \beta$.

Obviously, algorithm (15) with blocking scheme B-I is the same as the conventional GEE algorithm [cf., expression (14) in Appendix B]. Blocking scheme B-II divides all parameters into 2 blocks, while blocking scheme B-III divides the parameters into 4 blocks.

In (15) initial values of $(c_{0k}, c_{1k})$ are set to be their logistic regression estimates of model (4), for $k = 1, 2, 3$. Initial value of $\beta$ is set to be the average of the three logistic regression estimates of $\beta$ obtained from model (4) when $k = 1, 2, 3$. From 100 replicated simulations, Table 1 presents the averaged estimates of all parameters by algorithm (15) and the corresponding standard errors (in parentheses), when the three blocking schemes defined above are used. From the table, we can see that (i) parameter estimates with different blocking schemes are almost identical, and (ii) algorithm (15) estimates the parameters reasonably well.

3.2. *Real data analysis.* In this part we present some results about the tantrum data described in Section 2, using the modeling procedure (1)–(3). To estimate parameters in (2) and (3), the proposed modified version of the GEE algorithm (15) is used, after all parameters are grouped into the following 9 blocks: $(\mathbf{a}', \mathbf{b}')', \mathbf{c}_1, \ldots, \mathbf{c}_8$. From Table 1, we know that different blocking schemes would have minimal effect on the performance of (15).

To implement (15), we first need to choose a set of initial values $\hat{\tilde{\mathbf{c}}}^{(0)}$ for the parameters. While there are no existing general guidelines for this purpose in the GEE literature, we use the following procedure for choosing $\hat{\tilde{\mathbf{c}}}^{(0)}$. First, in (2), we let $\hat{a}_0^{(0)} = \log(1.5), \hat{b}_0^{(0)} = \log(3)$, and the remaining components of $\hat{\mathbf{a}}^{(0)}$ and $\hat{\mathbf{b}}^{(0)}$ be 0. Using these parameter values, the initial MA curve is assumed to be uncorrelated with duration $d$ and be skewed to the right with a peak at a quite early stage (cf., Figure 2), which is intuitively plausible. Second, in (3), we assume that all initial models are linear. Third, to choose initial values of $(\hat{c}_{0k}^{(0)}, \hat{c}_{1k}^{(0)})$, for $k = 1, 2, \ldots, 8$, we first obtain a set of empirical estimates $\hat{\pi}_k^{(\text{emp})}(t_j)$ of $\pi_k(t_j)$ at equally spaced time points $t_j = j \times 2/100$, for $k = 1, 2, \ldots, 8$ and $j = 1, 2, \ldots, 49$, where $\hat{\pi}_k^{(\text{emp})}(t_j)$ are defined to be relative frequencies of the $k$th angry behavior in the time intervals $[t_j - h, t_j + h]$, and $h = 0.1$ is a bandwidth. Then, $(\hat{c}_{0k}^{(0)}, \hat{c}_{1k}^{(0)})$ are chosen to be the estimated least squares coefficients of the fitted simple linear regression models from the datasets $\{(\text{MA}(t_j, \hat{\mathbf{a}}^{(0)}, \hat{\mathbf{b}}^{(0)}), \hat{\pi}_k^{(\text{emp})}(t_j)), j = 1, 2, \ldots, 49\}$, for $k = 1, 2, \ldots, 8$. By the way, we also tried several different values of $\hat{a}_0^{(0)}, \hat{b}_0^{(0)}$, and $h$; algorithm (15) gives similar results. The iterative algorithm (15) stops at the $j$th iteration when the relative difference between $\hat{\tilde{\mathbf{c}}}^{(j)}$ and $\hat{\tilde{\mathbf{c}}}^{(j-1)}$, defined as $\|\hat{\tilde{\mathbf{c}}}^{(j)} - \hat{\tilde{\mathbf{c}}}^{(j-1)}\| / \|\hat{\tilde{\mathbf{c}}}^{(j)}\|$, is smaller than or equal to 0.01, where $\| \cdot \|$ is the Euclidean norm.

To analyze the tantrum data using (1)–(3), we first need to determine the orders $m_a, m_b$ and $m_k$, for $k = 1, 2, \ldots, 8$, of the polynomial models (2) and (3). To this end, the "quasi-likelihood under the independence model criterion" (QIC) by Pan (2001) is used. To use this method, we can simply treat observations of the eight behavior variables of a given subject at a given time point as eight repeated measures of a single binary response. There are several slightly different versions of the QIC measure. Here, we use the version $QIC_u$ [see Pan (2001) for its definition] that is recommended by Pan (2001) and Cui and Qian (2007), due to its simplicity and its high success rates in selecting correct models in their numerical studies. To select a final model, we start from models (2)–(3) with $m_a = 2, m_b = 2$, and $m_k = 2$, for $k = 1, 2, \ldots, 8$, which is denoted as "Full" hereafter. Then, a backward model selection procedure is implemented, and some related results are summarized in Table 2. In the 'Model' columns of the table, we only list $m_a, m_b$ and $m_k$, for $k = 1, 2, \ldots, 8$, that are smaller than 2. So, for instance, model '$m_4 = m_1 = 1$' denotes the one with $m_a = 2; m_b = 2; m_1 = 1; m_2 = m_3 = 2; m_4 = 1; m_5 = m_6 = m_7 = m_8 = 2$.

From Table 2, it seems that models (2)–(3) with $m_4 = m_6 = m_a = 1$ and $m_1 = m_2 = m_3 = m_5 = m_7 = m_8 = m_b = 2$ fit the tantrum data best, by the $QIC_u$ criterion. So, we use these models in the remaining part of data analysis. By algorithm (15), the estimated parameter values and standard errors (SEs) of these models are presented in Table 3. The SEs are computed from the robust variance estimator $\hat{V}_R$ described in Appendix B.

TABLE 2
*This table lists some models and their $QIC_u$ values*

| Model | $QIC_u$ | Model | $QIC_u$ |
|---|---|---|---|
| Full | 12,274.49 | | |
| $m_1 = 1$ | 12,304.77 | $m_4 = m_1 = 1$ | 12,349.57 |
| $m_2 = 1$ | 12,326.92 | $m_4 = m_2 = 1$ | 12,325.00 |
| $m_3 = 1$ | 12,302.36 | $m_4 = m_3 = 1$ | 12,325.08 |
| $m_4 = 1$ | **12,272.03** | | |
| $m_5 = 1$ | 12,281.53 | $m_4 = m_5 = 1$ | 12,281.94 |
| $m_6 = 1$ | 12,274.67 | $m_4 = m_6 = 1$ | **12,272.00** |
| $m_7 = 1$ | 12,298.03 | $m_4 = m_7 = 1$ | 12,301.48 |
| $m_8 = 1$ | 12,276.45 | $m_4 = m_8 = 1$ | 12,275.12 |
| $m_4 = m_6 = m_1 = 1$ | 12,303.44 | $m_4 = m_6 = m_a = 1$ | **12,270.78** |
| $m_4 = m_6 = m_2 = 1$ | 12,325.18 | $m_4 = m_6 = m_b = 1$ | 12,328.93 |
| $m_4 = m_6 = m_3 = 1$ | 12,346.94 | $m_4 = m_6 = m_a = m_b = 1$ | 12,328.91 |
| | | $m_4 = m_6 = 1, m_a = 0$ | 12,330.26 |
| $m_4 = m_6 = m_5 = 1$ | 12,283.75 | | |
| $m_4 = m_6 = m_7 = 1$ | 12,305.91 | | |
| $m_4 = m_6 = m_8 = 1$ | 12,274.97 | | |

TABLE 3
*Estimated parameters of models (2) and (3) and their standard errors (in parentheses)*
*when $m_4 = m_6 = m_a = 1$ and $m_1 = m_2 = m_3 = m_5 = m_7 = m_8 = m_b = 2$*

| Parameters | Constant component | Linear component | Quadratic component |
|---|---|---|---|
| $\hat{\mathbf{a}}$ | 0.0658 (0.0479) | 0.0045 (0.0027) | — |
| $\hat{\mathbf{b}}$ | 0.2410 (0.0725) | 0.0454 (0.0100) | 0.0001 (0.0003) |
| $\hat{\mathbf{c}}_1$ | −6.1432 (0.6562) | 11.3835 (2.5709) | −8.6876 (2.8639) |
| $\hat{\mathbf{c}}_2$ | −1.6569 (0.1318) | −2.0083 (0.8088) | 3.4437 (1.1259) |
| $\hat{\mathbf{c}}_3$ | −1.5895 (0.1338) | 4.0299 (0.7408) | −4.0841 (1.1209) |
| $\hat{\mathbf{c}}_4$ | −5.0790 (0.3770) | 3.1491 (0.6375) | — |
| $\hat{\mathbf{c}}_5$ | −5.2077 (0.4210) | 5.8811 (1.8784) | −2.6364 (1.9953) |
| $\hat{\mathbf{c}}_6$ | −4.5346 (0.3031) | 3.3092 (0.5098) | — |
| $\hat{\mathbf{c}}_7$ | −4.2117 (0.3210) | 6.7421 (1.4772) | −4.9546 (1.7613) |
| $\hat{\mathbf{c}}_8$ | −3.8145 (0.3129) | −1.2724 (1.9000) | 2.8053 (2.3235) |

The estimated MA surface is presented in Figure 3(a). From the plot, it can be seen that (i) when duration $d$ is small, MA starts at a quite high level and then decreases relatively slowly over time $t$, (ii) as $d$ increases, MA starts at a lower level and drops faster over time, (iii) for a given $d$ value, MA peaks at an early time point, and (iv) it seems that the peak of MA decreases with duration $d$ when $d$ is small to moderate and then stabilizes when $d$ is large. To further demonstrate these results, in Figure 3(b), cross sections of MA when $d = 2, 5, 10, 20$ and $30$ are presented in a single plot, from which the above results can be easily seen. It should be mentioned that these results are all intuitively reasonable. For instance, in practice, an anger episode of a long duration usually starts at a relatively low intensity, the peak of its intensity is often at the early stage of the episode, and its intensity decreases quite fast in the entire episode. Plots (a) and (b) in Figure 3 are made with standardized time $t$ which represents the percentile of the duration of a tantrum episode at the corresponding real observation time, as discussed in Section 2. For convenience, to perceive the estimated MA function in real time, the estimated MA surface and its cross sections when $d = 2, 5, 10, 20$ and $30$ are presented again in Figures 3(c) and 3(d), respectively, in real time. It seems that all conclusions made from Figures 3(a) and 3(b) are still true here, except that the real peak time seems to increase with duration $d$ when $d$ is small and then stabilizes when $d$ gets larger.

The estimated probabilities of the eight angry behaviors are presented in Figure 4 as functions of MA, along with their pointwise 95% confidence intervals. The confidence intervals are computed based on the robust variance estimator $\hat{V}_R$. From the figure, it can be seen that (i) Stamp and Throw do not happen quite often and their likelihood of occurrence is almost flat over the entire range of anger intensity, (ii) Push and Hit do not happen quite often either, their likelihood of occurrence increases with MA, and both increasing trends look close to linear, (iii) the probabilities of Stiffen and Kick are at quite low levels in the range of MA, they first increase with MA and then stabilize, and (iv) Shout and Scream have relatively large probabilities of occurrence, the probability of Shout is quite stable when MA is below 0.4 and it increases with MA afterward, and the probability of Scream tends to increase with MA when MA is below 0.5 and decrease afterward.

To investigate goodness of fit of the estimated models shown in Figures 3 and 4, we first apply the Hosmer–Lemeshow test [cf., Hosmer and Lemeshow (1989)] to individual angry behaviors. By this approach, observations of the $k$th angry behavior are partitioned into $g$ groups using percentiles of the estimated probabilities. Here, we follow the convention that $g$ is chosen to be 10 and deciles of the estimated probabilities are used for grouping. Then, the Pearson's Chi-square statistic is defined by

$$X_k^2 = \sum_{\ell=1}^{10} \frac{(O_{k\ell} - E_{k\ell})^2}{E_{k\ell}} \qquad \text{for } k = 1, 2, \ldots, 8,$$

where

$$E_{k\ell} = \sum_{\pi_k(t_{ij}, \hat{\mathbf{a}}, \hat{\mathbf{b}}, \hat{\mathbf{c}}_k) \in (0.1(\ell-1), 0.1\ell]} \pi_k(t_{ij}, \hat{\mathbf{a}}, \hat{\mathbf{b}}, \hat{\mathbf{c}}_k),$$

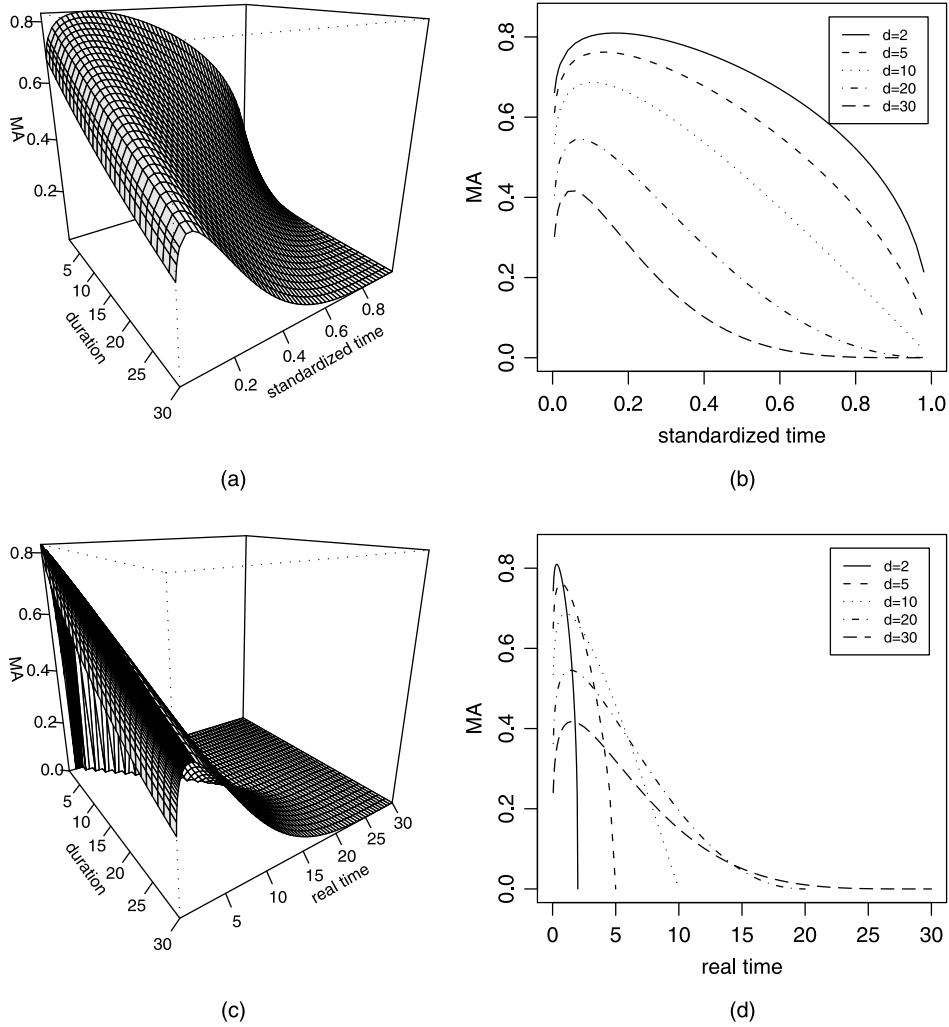

FIG. 3.   (a) *Estimated* MA *surface is shown in standardized time t.* (b) *Its cross sections when d = 2, 5, 10, 20 and 30 are shown by the solid, short-dashed, dotted, dot-dashed and long-dashed curves.* (c)–(d) *Same results as those in plots* (a)–(b) *are shown in real time.*

$$O_{k\ell} = \sum_{\pi_k(t_{ij}, \hat{\mathbf{a}}, \hat{\mathbf{b}}, \hat{\mathbf{c}}_k) \in (0.1(\ell-1), 0.1\ell]} y_{ijk}$$

are the expected and observed counts of the $\ell$th group, respectively, and $y_{ijk}$ is the observed $k$th angry behavior at time $t_{ij}$. In the above expressions, when $\ell = 1$, the grouping interval $(0, 0.1]$ can be changed to $[0, 0.1]$, although this change would not make much difference in the results because the estimated probabilities would not be exactly 0 in most cases. By these formulas, the calculated values of $X_k^2$ are listed in Table 4.

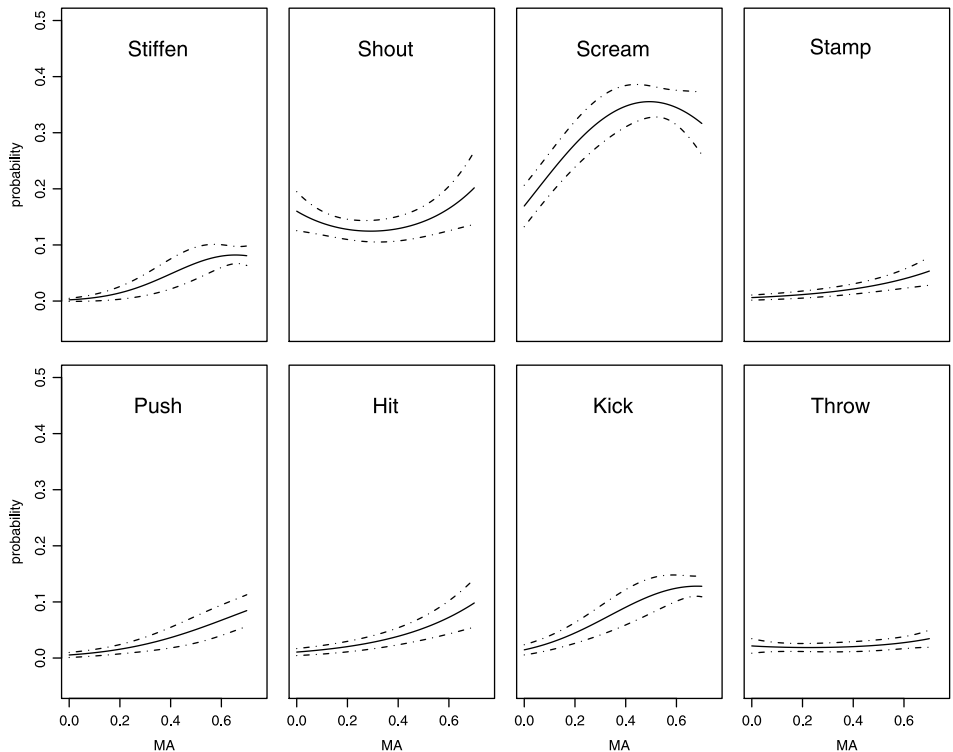

FIG. 4. *Estimated probabilities of the eight angry behaviors as functions of* MA, *along with the pointwise 95% confidence intervals for the probabilities.*

TABLE 4
*Calculated values of $X_k^2$, for $k = 1, 2, \ldots, 8$*

| $k$ | 1 | 2 | 3 | 4 |
|---|---|---|---|---|
| $X_k^2$ | $2.03 \times 10^{-6}$ | $2.85 \times 10^{-4}$ | $1.16$ | $1.63 \times 10^{-6}$ |
| $k$ | 5 | 6 | 7 | 8 |
| $X_k^2$ | $1.53 \times 10^{-2}$ | $1.38 \times 10^{-1}$ | $1.59 \times 10^{-1}$ | $4.76 \times 10^{-7}$ |

According to Hosmer and Lemeshow (1989), if the chosen model fits the data well, then $X_k^2$ should follow a Chi-square distribution with 8 degrees of freedom. The 95th percentile of this distribution is 15.51, which is much larger than all $X_k^2$ values in Table 4. Therefore, the estimated model fits the data well in terms of individual angry behaviors. If we would like to study how well the estimated models fit the entire dataset, then a reasonable test

statistic is

$$X^2 = \sum_{k=1}^{8} X_k^2.$$

In the two extreme cases that $\{X_k, k = 1, 2, \ldots, 8\}$ are independent of each other and that they are perfectly linearly correlated, the null distribution of $X^2$ would be $\chi^2(64)$ and $8\chi^2(8)$, respectively, with 95th percentiles 83.68 and 124.06. By Table 4, the calculated value of $X^2$ from the observed data is 1.48, which is much smaller than either one of the two 95th percentiles. So, by the Hosmer–Lemeshow test, we can conclude that the estimated models shown in Figures 3 and 4 fit the observed data well.

To further check the adequacy of the estimated models, next we use a graphical approach by making certain model checking plots, introduced in detail in Chapter 22 of Cook and Weisberg (1999). The basic idea of a model checking plot is to present the predicted response values based on the proposed model and the corresponding empirically estimated response values without using the proposed model in a single plot. If the two sets of values are close to each other, then we conclude that the proposed model describes the observed data well. Otherwise, the proposed model may not be appropriate to use. Although this model checking approach is based on our visual impression and may not be mathematically rigorous, it is a useful tool, especially when related statistical theory is not available. For the tantrum data, we can compare the predicted probabilities of the eight angry behaviors based on models (1)–(3) and the corresponding empirical estimates of these probabilities computed directly from the observed data. Since (i) models (1)–(3) assume that these probabilities depend on both duration $d$ and time $t$, (ii) their empirical estimates can be computed at a given time point, and (iii) the empirical estimates can not be computed at a given MA level due to the fact that MA is unobservable, we choose to present predicted probabilities based on models (1)–(3) and the empirical estimates of the probabilities when $d = 2$ or 8 minutes and $t = j \times 0.05$, for $j = 1, 2, \ldots, 19$, in the model checking plots. The two specific duration values are chosen based on the following considerations. First, because most durations in the data are quite small (cf., Figure 1), we can not choose large duration values. Otherwise, both empirical probability estimates and predicted probabilities based on models (1)–(3) would have large variability. Second, the two chosen duration values should be different enough to demonstrate how probabilities of angry behaviors depend on duration. The predicted probabilities can be computed from the estimated models of (2) and (3). For given values of $d$ and $t$, the empirical probability estimates are taken to be the sample proportions of occurrence of the angry behaviors in the time interval $[t - 0.05, t + 0.05]$, duration interval $[d - 1, d + 1]$ when $d = 2$, and duration interval $[d - 3, d + 3]$ when $d = 8$. The model checking plots for the eight angry behaviors are shown in the upper eight panels of Figure 5 when $d = 2$, and in the lower eight panels of Figure 5 when $d = 8$. From the plots, it can be seen that overall the predicted probabilities and the empirical estimates of the probabilities match reasonably well. For the behavior Scream, the predicted probabilities overestimate the trend over time of the empirical probabilities a little bit when $d = 2$ and they match well when $d = 8$. Small mismatches can also be noticed in several other places (e.g., for Hit when $d = 2$, and for Shout when $d = 8$).

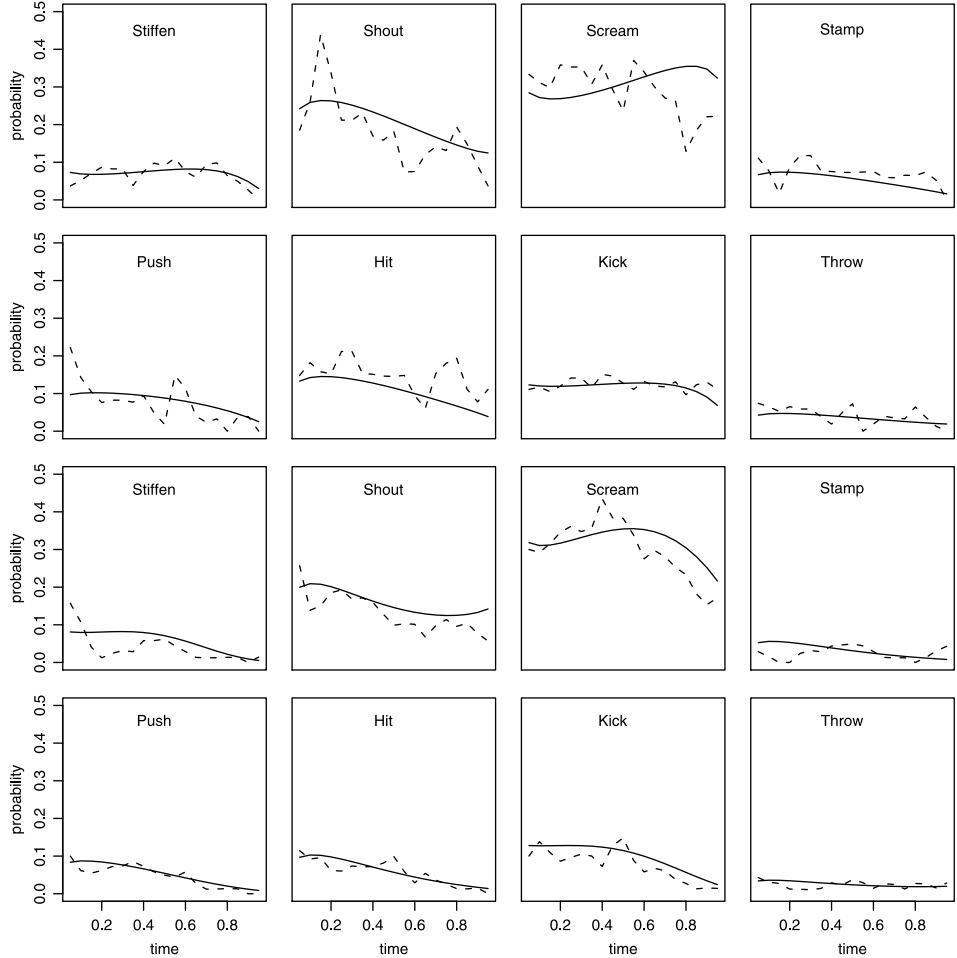

FIG. 5. *The upper eight panels show the model checking plots of the proposed models (1)–(3) when $d = 2$. The lower eight panels show the model checking plots when $d = 8$.*

**4. Summary and concluding remarks.** We have presented a procedure for modeling both the time course of tantrum anger and the relationship between anger and various behaviors that express it. In this model, each of eight angry behaviors is assumed to be uniquely driven by a latent variable MA that represents momentary anger intensity. MA is modeled by a Beta function over the time course of the tantrum. Its two parameters are assumed to be polynomial functions of tantrum duration $d$, and the linkage between MA and the angry behaviors is described by generalized polynomial models. Model parameters are estimated by a modified version of the GEE iterative algorithm, after model selection using the QIC criterion. Goodness of fit of the estimated models is checked by the Hosmer–Lemeshow test. Based on this analysis, we can make the following conclusions: (i) The selected models are appropriate for describing the tantrum data. (ii) The MA function peaks near the beginning of the tantrum for a given value of $d$ and then decreases gradually. This finding is consistent with the result of "rapid rise and slow decline of anger" reported in the psychological literature [cf., Beck and Fernandez (1998), Fridja et al. (1991), Tsytsarev and Grodnitzky (1995)]. (iii) The peak value of MA appears to decrease with tantrum duration $d$, in terms of the standardized time $t$. This interesting result was also seen in an earlier analysis of a different sort [cf., Potegal, Kosorok and Davidson (1996)]. (iv) The probability of the lower angry behaviors Stamp and Throw increases only slightly with MA. In contrast, the probability of higher angry behaviors like Shout, Scream, Hit and Kick changes significantly with MA. This finding provides an explanation for their differential distribution across the tantrum. Namely, the lower angry behaviors are broadly distributed because their probability is largely unaffected by changes in anger intensity, while the higher anger behaviors tend to appear around the peak of MA. At a more fundamental level, this model explains how various behaviors actually come to reflect different levels of anger intensity.

To the best of our knowledge, this analysis is the first to reconstruct the time course of anger based on objective behavioral data. It also proposes an approach to the analysis of multiple, binary, longitudinal (MBL) data with a latent variable involved. Such data are quite common in psychological, psychiatric and other areas of social sciences. From a psychological perspective, one innovative aspect of the proposed approach is that it starts with the temporal relations among various objectively observed behaviors in a naturally occurring situation and then works backward to determine the corresponding levels of anger intensity, rather than starting with subjective estimates of anger intensity and trying to determine properties of the associated behaviors.

At the end of the article, we would like to point out that, besides the proposed modeling approach and the corresponding GEE model estimation algorithm, there might be other ways to analyze MBL data. One possibility is to use mixed-effects modeling [e.g., Diggle (1988), Laird and Ware (1982), Lindstrom and Bates (1988)]. By introducing a random-effects term for describing between-subject variability, correlation among repeated measures within a subject can be accommodated to a certain degree. It requires much future research to set up such a random-effects model in an appropriate way, and to compare this alternative approach to the approach discussed in this paper. We adopt the GEE approach here because (i) correlation among observations does not have to be precisely specified by this approach and estimates are consistent under regularity conditions, and (ii) from a technical or mathematical standpoint, it is easier to incorporate the latent variable MA in that framework. After the parameters in models (2) and (3) are estimated, it might be of interest to some researchers to compare the effect of MA on the eight different angry behaviors. To this end, an appropriate multiple comparison procedure should be applied, which also requires much future research.

The tantrum data analyzed in this paper and the R code fitting the final model selected by QIC and presented in Figures 3 and 4 and Table 3 are available online as supplementary materials [Qiu, Yang and Potegal (2009)].

## APPENDIX A: PARAMETER IDENTIFIABILITY IN MODELS (1)–(3)

THEOREM A.1. *In models (1)–(3), all parameters $\mathbf{a}, \mathbf{b}$ and $\mathbf{c}_k$, for $k = 1, 2, \ldots, 8$, are identifiable.*

PROOF. To verify the identifiability of parameters in models (1)–(3), we need to show that, if there are two sets of parameters $\{\mathbf{a}^{(1)}, \mathbf{b}^{(1)}, \mathbf{c}_k^{(1)}, k = 1, 2, \ldots, 8\}$ and $\{\mathbf{a}^{(2)}, \mathbf{b}^{(2)}, \mathbf{c}_k^{(2)}, k = 1, 2, \ldots, 8\}$ such that, for all $t \in [0, 1]$,

$$\begin{aligned}
(5) \qquad &\mathrm{logit}(\pi_k(t, \mathbf{a}^{(1)}, \mathbf{b}^{(1)}, \mathbf{c}_k^{(1)})) \\
&= \mathrm{logit}(\pi_k(t, \mathbf{a}^{(2)}, \mathbf{b}^{(2)}, \mathbf{c}_k^{(2)})) \qquad \text{for } k = 1, 2, \ldots, 8,
\end{aligned}$$

then we must have $\mathbf{a}^{(1)} = \mathbf{a}^{(2)}, \mathbf{b}^{(1)} = \mathbf{b}^{(2)}$, and $\mathbf{c}_k^{(1)} = \mathbf{c}_k^{(2)}$, for $k = 1, 2, \ldots, 8$.

For $l = 1, 2$, let $\mathbf{a}^{(l)} = (a_0^{(l)}, a_1^{(l)}, \ldots, a_{m_a}^{(l)})', \mathbf{b}^{(l)} = (b_0^{(l)}, b_1^{(l)}, \ldots, b_{m_b}^{(l)})', \mathbf{c}_k^{(l)} = (c_{0k}^{(l)}, c_{1k}^{(l)}, \ldots, c_{m_k k}^{(l)})'$, and

$$a^{(l)} = a_0^{(l)} + a_1^{(l)} d + \cdots + a_{m_a}^{(l)} d^{m_a}, \qquad b^{(l)} = b_0^{(l)} + b_1^{(l)} d + \cdots + b_{m_b}^{(l)} d^{m_b}.$$

Equation (5) implies that, for $t \in [0, 1]$,

$$\begin{aligned}
(6) \qquad &c_{0k}^{(1)} + c_{1k}^{(1)} \mathrm{MA}(t, a^{(1)}, b^{(1)}) + \cdots + c_{m_k k}^{(1)} \mathrm{MA}^{m_k}(t, a^{(1)}, b^{(1)}) \\
&= c_{0k}^{(2)} + c_{1k}^{(2)} \mathrm{MA}(t, a^{(2)}, b^{(2)}) + \cdots + c_{m_k k}^{(2)} \mathrm{MA}^{m_k}(t, a^{(2)}, b^{(2)}).
\end{aligned}$$

Obviously, $\mathrm{MA}(t, a^{(l)}, b^{(l)})$ can be written as a polynomial function of $t$. For a given $d$ value, without loss of generality, assume that $a^{(l)}$ and $b^{(l)}$ are positive for $l = 1, 2$. Then, the constant terms of the two polynomial functions on the two different sides of (6) are $c_{0k}^{(1)}$ and $c_{0k}^{(2)}$, respectively. So, we have

$$(7) \qquad c_{0k}^{(1)} = c_{0k}^{(2)} \qquad \text{for } k = 1, 2, \ldots, 8.$$

Equations (6) and (7) imply that, for $t \in [0, 1]$,

$$(8) \quad \begin{aligned} c_{1k}^{(1)} \, \mathrm{MA}(t, a^{(1)}, b^{(1)}) + \cdots + c_{m_k k}^{(1)} \, \mathrm{MA}^{m_k}(t, a^{(1)}, b^{(1)}) \\ = c_{1k}^{(2)} \, \mathrm{MA}(t, a^{(2)}, b^{(2)}) + \cdots + c_{m_k k}^{(2)} \, \mathrm{MA}^{m_k}(t, a^{(2)}, b^{(2)}). \end{aligned}$$

The lowest-order terms of the two polynomial functions on the two sides of (8) are $c_{1k}^{(1)} t^{e^{a^{(1)}} - 1}$ and $c_{1k}^{(2)} t^{e^{a^{(2)}} - 1}$, respectively. Therefore, they must be the same. Namely,

$$c_{1k}^{(1)} = c_{1k}^{(2)} \qquad \text{for } k = 1, 2, \ldots, 8$$

and

$$(9) \qquad a^{(1)} = a^{(2)} \qquad \text{for all } d.$$

Using such arguments recursively, we have

$$\mathbf{c}_k^{(1)} = \mathbf{c}_k^{(2)} \qquad \text{for } k = 1, 2, \ldots, 8.$$

From (9), we have

$$\mathbf{a}^{(1)} = \mathbf{a}^{(2)}.$$

By comparing the highest-order terms of the two polynomial functions on the two sides of (8), we have

$$(10) \qquad b^{(1)} = b^{(2)} \qquad \text{for all } d.$$

From (10), we have

$$\mathbf{b}^{(1)} = \mathbf{b}^{(2)}.$$

Therefore, the identifiability of parameters in models (1)–(3) is proved. $\quad\square$

## APPENDIX B: MODEL ESTIMATION

In this part we discuss estimation of the parameters in models (1)–(3) from the tantrum data under the framework of generalized estimating equations (GEE). By this approach, as long as the mean function of observations is correctly specified and their variance structure is roughly specified, unknown parameters in the models can be estimated by a GEE iterative algorithm [cf., Liang and Zeger ([1986])].

Let $y_{ijk}$ denote the binary observation of the $k$th angry behavior of the $i$th child at the time point $t_{ij}$, for $j = 1, 2, \ldots, n_i$, $i = 1, 2, \ldots, N$, and $k = 1, 2, \ldots, 8$, with $y_{ijk} = 1$ denoting presence of the behavior and 0 absence. Let $\mathbf{Y}_{ij} = (y_{ij1}, y_{ij2}, \ldots, y_{ij8})'$ be the vector of eight angry behaviors observed at the time point $t_{ij}$ for the $i$th child. Its mean vector can be written as

$$(11) \quad \boldsymbol{\pi}(t_{ij}, \tilde{\mathbf{c}}) = (\pi_1(t_{ij}, \mathbf{a}, \mathbf{b}, \mathbf{c}_1), \pi_2(t_{ij}, \mathbf{a}, \mathbf{b}, \mathbf{c}_2), \ldots, \pi_8(t_{ij}, \mathbf{a}, \mathbf{b}, \mathbf{c}_8))',$$

where $\tilde{\mathbf{c}} = (\mathbf{a}', \mathbf{b}', \mathbf{c}')'$ and $\mathbf{c} = (\mathbf{c}_1', \mathbf{c}_2', \ldots, \mathbf{c}_8')'$. The covariance matrix of $\mathbf{Y}_{ij}$ can be written as

$$(12) \quad V(t_{ij}, \tilde{\mathbf{c}}, \alpha) = A^{1/2}(t_{ij}, \tilde{\mathbf{c}}) R(t_{ij}, \alpha) A^{1/2}(t_{ij}, \tilde{\mathbf{c}}),$$

where $R(t_{ij}, \alpha)$ is a "working" correlation matrix of $\mathbf{Y}_{ij}$ which may depend on a parameter (vector) $\alpha$, and

$$A(t_{ij}, \tilde{\mathbf{c}}) = \text{diag}(\pi_1(t_{ij}, \mathbf{a}, \mathbf{b}, \mathbf{c}_1)(1 - \pi_1(t_{ij}, \mathbf{a}, \mathbf{b}, \mathbf{c}_1)),$$
$$\pi_2(t_{ij}, \mathbf{a}, \mathbf{b}, \mathbf{c}_2)(1 - \pi_2(t_{ij}, \mathbf{a}, \mathbf{b}, \mathbf{c}_2)),$$
$$\ldots, \pi_8(t_{ij}, \mathbf{a}, \mathbf{b}, \mathbf{c}_8)(1 - \pi_8(t_{ij}, \mathbf{a}, \mathbf{b}, \mathbf{c}_8))).$$

Then, the generalized estimating equations are defined by

$$(13) \quad \sum_{i=1}^{N} \sum_{j=1}^{n_i} \left( \frac{\partial \boldsymbol{\pi}(t_{ij}, \tilde{\mathbf{c}})}{\partial \tilde{\mathbf{c}}} \right)' V^{-1}(t_{ij}, \tilde{\mathbf{c}}, \alpha)(\mathbf{Y}_{ij} - \boldsymbol{\pi}(t_{ij}, \tilde{\mathbf{c}})) = \mathbf{0}.$$

By Liang and Zeger's ([1986]) approach, the estimator of $\tilde{\mathbf{c}}$ can be computed by iterating between the following Fisher scoring algorithm for $\tilde{\mathbf{c}}$ and a moment estimation for $\alpha$. Let $\hat{\tilde{\mathbf{c}}}^{(0)}$ be an initial estimator of $\tilde{\mathbf{c}}$. Then, in the $j$th iteration, for any $j > 0$, the updated estimator of $\tilde{\mathbf{c}}$ is defined by

$$\hat{\tilde{\mathbf{c}}}^{(j)} = \hat{\tilde{\mathbf{c}}}^{(j-1)}$$

$$+ \left[ \sum_{i=1}^{N} \sum_{j=1}^{n_i} \left( \frac{\partial \boldsymbol{\pi}(t_{ij}, \hat{\tilde{\mathbf{c}}}^{(j-1)})}{\partial \tilde{\mathbf{c}}} \right)' \right.$$

$$(14) \qquad \left. \times V^{-1}(t_{ij}, \hat{\tilde{\mathbf{c}}}^{(j-1)}, \hat{\alpha}(\hat{\tilde{\mathbf{c}}}^{(j-1)})) \left( \frac{\partial \boldsymbol{\pi}(t_{ij}, \hat{\tilde{\mathbf{c}}}^{(j-1)})}{\partial \tilde{\mathbf{c}}} \right) \right]^{-1}$$

$$\times \left[ \sum_{i=1}^{N} \sum_{j=1}^{n_i} \left( \frac{\partial \boldsymbol{\pi}(t_{ij}, \hat{\tilde{\mathbf{c}}}^{(j-1)})}{\partial \tilde{\mathbf{c}}} \right)' \right.$$

$$\left. \times V^{-1}(t_{ij}, \hat{\tilde{\mathbf{c}}}^{(j-1)}, \hat{\alpha}(\hat{\tilde{\mathbf{c}}}^{(j-1)})) (\mathbf{Y}_{ij} - \boldsymbol{\pi}(t_{ij}, \hat{\tilde{\mathbf{c}}}^{(j-1)})) \right],$$

where $\hat{\alpha}(\hat{\tilde{\mathbf{c}}}^{(j-1)})$ is the estimator of $\alpha$ when $\tilde{\mathbf{c}}$ is estimated by $\hat{\tilde{\mathbf{c}}}^{(j-1)}$.

The "working" correlation matrix can be modeled in several different ways. These include the identity matrix $I_{8\times 8}$ (when it is reasonable to assume that the eight angry behavior variables are all independent of each other), the compound symmetry structure or, equivalently, the exchangeable structure (when it is assumed that correlations between any two angry behaviors are all the same), the unstructured pattern where we need to estimate all $8(8-1)/2$ off-diagonal elements of $R(t_{ij}, \alpha)$ and so forth. After the pattern of $R(t_{ij}, \alpha)$ is determined, $\alpha$ can be estimated using the method of moments. For instance, if $R(t_{ij}, \alpha)$ is assumed to have the compound symmetry structure with its $(k_1, k_2)$th element being

$$R_{k_1, k_2}(t_{ij}, \alpha) = \begin{cases} 1, & \text{if } k_1 = k_2, \\ \alpha, & \text{otherwise,} \end{cases}$$

then $\alpha$ is the correlation coefficient between any two angry behaviors and its moment estimator is given by

$$\hat{\alpha}(\hat{\tilde{\mathbf{c}}}^{(j-1)}) = \frac{1}{N^* - p} \sum_{i=1}^{N} \sum_{j=1}^{n_i} \sum_{k_1 < k_2} r_{ijk_1} r_{ijk_2},$$

where $N^* = \sum_{i=1}^{N} \sum_{j=1}^{n_i} (8 \times 7)/2$, $p$ is the number of parameters in the mean function [cf., equation (11)], $r_{ijk} = (y_{ijk} - \boldsymbol{\pi}(t_{ij}, \hat{\tilde{\mathbf{c}}}^{(j-1)}))/[\boldsymbol{\pi}(t_{ij}, \hat{\tilde{\mathbf{c}}}^{(j-1)})(1 - \boldsymbol{\pi}(t_{ij}, \hat{\tilde{\mathbf{c}}}^{(j-1)}))]^{1/2}$, and $\hat{\tilde{\mathbf{c}}}^{(j-1)}$ is the estimated parameter vector obtained from the GEE algorithm [cf., equation (14)]. Obviously, in the above expression, $\{r_{ijk}\}$ are the standardized residuals and $\hat{\alpha}(\hat{\tilde{\mathbf{c}}}^{(j-1)})$ is the pooled sample correlation constructed from these residuals.

Let $\hat{\tilde{\mathbf{c}}}$ be the solution of (13). According to Liang and Zeger (1986), under some mild regularity conditions, $N^{1/2}(\hat{\tilde{\mathbf{c}}} - \tilde{\mathbf{c}})$ is asymptotically Normal with mean zero and covariance matrix

$$V_R = \lim_{N \to \infty} \left( \sum_{i=1}^{N} \sum_{j=1}^{n_i} \left( \frac{\partial \boldsymbol{\pi}(t_{ij}, \tilde{\mathbf{c}})}{\partial \tilde{\mathbf{c}}} \right)' V^{-1}(t_{ij}, \tilde{\mathbf{c}}, \alpha) \left( \frac{\partial \boldsymbol{\pi}(t_{ij}, \tilde{\mathbf{c}})}{\partial \tilde{\mathbf{c}}} \right) \right)^{-1}$$

$$\times \left( \sum_{i=1}^{N} \sum_{j=1}^{n_i} \left( \frac{\partial \boldsymbol{\pi}(t_{ij}, \tilde{\mathbf{c}})}{\partial \tilde{\mathbf{c}}} \right)' V^{-1}(t_{ij}, \tilde{\mathbf{c}}, \alpha) \right.$$

$$\times \operatorname{Cov}(\mathbf{Y}_{ij})V^{-1}(t_{ij}, \tilde{\mathbf{c}}, \alpha)\left(\frac{\partial \boldsymbol{\pi}(t_{ij}, \tilde{\mathbf{c}})}{\partial \tilde{\mathbf{c}}}\right)\right)$$

$$\times \left(\sum_{i=1}^{N}\sum_{j=1}^{n_i}\left(\frac{\partial \boldsymbol{\pi}(t_{ij}, \tilde{\mathbf{c}})}{\partial \tilde{\mathbf{c}}}\right)' V^{-1}(t_{ij}, \tilde{\mathbf{c}}, \alpha)\left(\frac{\partial \boldsymbol{\pi}(t_{ij}, \tilde{\mathbf{c}})}{\partial \tilde{\mathbf{c}}}\right)\right)^{-1},$$

where $\operatorname{Cov}(\mathbf{Y}_{ij})$ denotes the covariance matrix of $\mathbf{Y}_{ij}$. In applications, $V_R$ can be estimated by its finite-sample version after $\tilde{\mathbf{c}}$ and $\alpha$ are replaced respectively by $\hat{\tilde{\mathbf{c}}}$ and $\hat{\alpha}$, and $\operatorname{Cov}(\mathbf{Y}_{ij})$ by $(\mathbf{Y}_{ij} - \boldsymbol{\pi}(t_{ij}, \hat{\tilde{\mathbf{c}}}))(\mathbf{Y}_{ij} - \boldsymbol{\pi}(t_{ij}, \hat{\tilde{\mathbf{c}}}))'$. The resulting estimator $\hat{V}_R$ is often called the "robust" variance estimator.

A favorable property of the GEE method is that the above-mentioned asymptotic normality of $\hat{\tilde{\mathbf{c}}}$ depends only on the correct specification of the mean function [cf., (11)]; it does not depend on the correct choice of the "working" correlation matrix $R(t_{ij}, \alpha)$ in (12). For this reason, if we do not have any prior information about the correlation matrix of $\mathbf{Y}_{ij}$, then $R(t_{ij}, \alpha)$ is often chosen to be the identity matrix in applications [cf., Diggle et al. (1994), Section 8.4.2].

In models (11) and (12), besides $\alpha$, there are 30 parameters in the parameter vector $\tilde{\mathbf{c}}$. In the updating formula (14) of the conventional GEE algorithm, we need to compute the inverse of the $30 \times 30$ matrix

$$\sum_{i=1}^{N}\sum_{j=1}^{n_i}(\partial \boldsymbol{\pi}(t_{ij}, \hat{\tilde{\mathbf{c}}}^{(j-1)})/\partial \tilde{\mathbf{c}})' V^{-1}(t_{ij}, \hat{\tilde{\mathbf{c}}}^{(j-1)}, \hat{\alpha}(\hat{\tilde{\mathbf{c}}}^{(j-1)}))(\partial \boldsymbol{\pi}(t_{ij}, \hat{\tilde{\mathbf{c}}}^{(j-1)})/\partial \tilde{\mathbf{c}})$$

in each iteration, which requires a great amount of CPU time. To reduce the computing burden, we suggest dividing the vector $\tilde{\mathbf{c}}$ into $g$ blocks, that is, $\tilde{\mathbf{c}} = (\tilde{\mathbf{c}}_1', \tilde{\mathbf{c}}_2', \ldots, \tilde{\mathbf{c}}_g')'$, and then using the following Fisher scoring algorithm with blocked parameters. In the $j$th iteration, for any $j > 0$, update the estimator of $\tilde{\mathbf{c}}_\ell$, for $\ell = 1, 2, \ldots, g$, successively by the updating formula

$$
\begin{aligned}
\hat{\tilde{\mathbf{c}}}_\ell^{(j)} &= \hat{\tilde{\mathbf{c}}}_\ell^{(j-1)} \\
(15) \quad &+ \left[\sum_{i=1}^{N}\sum_{j=1}^{n_i}\left(\frac{\partial \boldsymbol{\pi}(t_{ij}, \hat{\tilde{\mathbf{c}}}^*)}{\partial \tilde{\mathbf{c}}_\ell}\right)' V^{-1}(t_{ij}, \hat{\tilde{\mathbf{c}}}^*, \hat{\alpha}(\hat{\tilde{\mathbf{c}}}^*))\left(\frac{\partial \boldsymbol{\pi}(t_{ij}, \hat{\tilde{\mathbf{c}}}^*)}{\partial \tilde{\mathbf{c}}_\ell}\right)\right]^{-1} \\
&\times \left[\sum_{i=1}^{N}\sum_{j=1}^{n_i}\left(\frac{\partial \boldsymbol{\pi}(t_{ij}, \hat{\tilde{\mathbf{c}}}^*)}{\partial \tilde{\mathbf{c}}_\ell}\right)' V^{-1}(t_{ij}, \hat{\tilde{\mathbf{c}}}^*, \hat{\alpha}(\hat{\tilde{\mathbf{c}}}^*))(\mathbf{Y}_{ij} - \boldsymbol{\pi}(t_{ij}, \hat{\tilde{\mathbf{c}}}^*))\right],
\end{aligned}
$$

where $\hat{\tilde{\mathbf{c}}}^* = (\hat{\tilde{\mathbf{c}}}_1^{(j)}, \ldots, \hat{\tilde{\mathbf{c}}}_{\ell-1}^{(j)}, \hat{\tilde{\mathbf{c}}}_\ell^{(j-1)}, \ldots, \hat{\tilde{\mathbf{c}}}_g^{(j-1)})'$.

**Acknowledgments.** The authors thank the AE and two referees for many constructive comments and suggestions which greatly improved the quality of the paper.

## SUPPLEMENTARY MATERIAL

**Tantrum data and R Code** (DOI: [10.1214/09-AOAS242SUPP](); .zip). This is the tantrum anger data analyzed in the paper. The data has 10 columns. The first 8 columns denote the binary status of the 8 angry behaviors, with 1 denoting "present" and 0 "absent." The 9th column is the duration of a tantrum episode, and the 10th column is the standardized observation time. The data are ordered by duration (i.e., the 9th column). This is a R code fitting the final model selected by QIC presented in Figures 3 and 4 and Table 3 of the paper.

P. Qiu
School of Statistics
University of Minnesota
313 Ford Hall, 224 Church St. SE
Minneapolis, Minnesota 55455
USA
E-mail: qiu@stat.umn.edu

R. Yang
Pharmaceutical Research Institute
Bristol-Myers Squibb
211 Pomeroy Ave.
Meriden, Connecticut 06450
USA
E-mail: rong.yang@bms.com

M. Potegal
Department of Pediatrics
University of Minnesota
420 Delaware St. SE
Minneapolis, Minnesota 55455
USA
E-mail: poteg001@umn.edu